\documentclass[twocolumn]{aastex62}
\received{March 7, 2023}
\revised{May 8, 2023}
\accepted{May 28, 2023}
\usepackage{xcolor}
\usepackage{graphicx}
\usepackage{natbib}
\usepackage{subfigure}
\usepackage{float}
\usepackage{amsmath}
\usepackage{color}
\usepackage{graphicx}
\usepackage{amssymb}
\usepackage{amsmath}
\graphicspath{{figures/}}
\usepackage{enumitem}
\begin{document}
   \title{The effects of rotation, metallicity and magnetic field on the islands of failed supernovae}
\correspondingauthor{Chunhua Zhu,Guoliang L\"{u}}
\email{chunhuazhu@sina.cn,guolianglv@xao.ac.cn}
\author{Lei Li}
\affil{School of Physical Science and Technology,Xinjiang University, Urumqi, 830046, China}
\affil{ Xinjiang Astronomical Observatory, Chinese Academy of Sciences, 150 Science 1-Street, Urumqi, Xinjiang 830011, China}
\author{Chunhua Zhu*}
\affil{School of Physical Science and Technology,Xinjiang University, Urumqi, 830046, China}
\author{Sufen Guo}
\affil{School of Physical Science and Technology,Xinjiang University, Urumqi, 830046, China}
\author{Helei Liu}
\affil{School of Physical Science and Technology,Xinjiang University, Urumqi, 830046, China}
\author{Guoliang L\"{u}*}
\affil{ Xinjiang Astronomical Observatory, Chinese Academy of Sciences, 150 Science 1-Street, Urumqi, Xinjiang 830011, China}
\affil{School of Physical Science and Technology,Xinjiang University, Urumqi, 830046, China}

\begin{abstract}
   Failed supernovae (FSN) are a possible channel for the formation of heavy stellar-mass black holes ($M_{ BH}>\sim 30$ M$_\odot$). However, the effects of metallicity, rotation and magnetic field on the islands of explodabilty of massive stars are not clear. Here, we simulate the stellar structure and evolution in the mass range between 6 and 55 $M_{\odot}$ with different initial rotational velocities, metallicities, and magnetic fields from zero-age main sequence (ZAMS) to pre-collapse. We find that the rapid rotating stars can remain lower $\rm ^{12}C$ mass fraction at the time of C ignition, which allows the transition, from convective carbon burning to radiative burning, to occur at lower $M_{\rm ZAMS}$ than those from stars without rotation. However, the rapid rotation is unfavorable for FSN occurring but is conducive to long gamma-ray bursts (lGRBs) because it results in the specific angular momentum in the CO core is greater than the last stable orbit at core collapse. The increasing metallicity does not affect FSN islands, but high metallicity inhibits rotational mixing and is unfavorable for producing lGRBs. A magnetic field can constrain the mass-loss rate even for rapid rotating stars, resulting in higher mass at pre-collapse. The magnetic braking triggered by the magnetic field can reduce the rotation velocity for high-metallicity models, which decreases the specific angular momentum in the CO core and is favorable for FSN occurring. We suggest that the heavy-mass black holes detected by LIGO may originate from rapidly rotating massive stars with strong magnetic fields, rather than those with very low metallicity.
\end{abstract}
%

\section{Introduction}
Massive stars, with masses above $\sim$ 8 $M_{\odot}$, experience core collapse at the end of their evolution. This can lead to a supernovae (SNe) explosion.
However, some massive stars do not seem to explode successfully and disappear after the iron core collapse without bright transient. Many studies have revealed that some stars are too massive to overcome their own gravity to produce a typical SN, but they collapse directly into black holes and devour all the material they cannot eject \citep[e.g.][]{Bodenheimer1983,Fryer1999,Fryer2000,Connor2011,Janka2012,Smartt2015}.

Population synthesis studies found that the red supergiant (RSG) is the progenitor of type \uppercase\expandafter{\romannumeral2}P SN \citep{Dyk2003}. \citet{Smartt2009} used the observational data of 132 SNe in 10 yr to carry out a statistical study. They found that none of the explosions of type IIP SN came from the RSG with a mass between about 16 and 25 $M_\odot$. This is the so-called RSG problem.
\cite{Horiuchi2011} compared the SNe explosion rate with the birth rate of massive stars, and found that the former is less than the latter,  whether in the Local Cluster or in the galaxies with high redshift. \citet{Smartt2009} suggested that these massive RSGs should collapse directly into black holes after experiencing failed supernovae (FSN). In addition, the upper mass limit of RSGs has been thoroughly discussed in recent years \citep{Davies2018,Kochanek2020}, which further confirms the existence of FSN.

\cite{Kochanek2008} first proposed a survey for FSN. Using the 4 yr survey data of Large Binocular Telescope,  they discovered the first candidate, N6946-BH1 \citep{Gerke2015,Basinger2021}. It was a red source consistent with a mass 18 -- 25 $M_\odot$ RSG star, which disappeared after a long-duration weak transient.
The light curve of N6946-BH1 is similar with theoretical predictions by \cite{Lovegrove2013}. \cite{Adams2017} confirmed that the candidate object is a disappeared massive object subsequently. \cite{Reynolds2015} reported a yellow supergiant with mass of 20 -- 30 $M_\odot$ that disappeared without a record neutron star. There is another luminous blue variable as a candidate of FSN \citep{Allan2020}. \cite{Neustadt2021} estimated the rate of FSN is $f$ = $0.16^{+0.23}_{-0.12}$ for all core-collapse SN. Recently, \cite{Byrne2022} suggested $f <$ 0.23 for sources with absolute magnitude lower than -14.

The gravitational waves observed by Advanced LIGO $\&$ Virgo indicate that there are a large number of binary black hole merger events in the universe, and many of them are heavier than the most massive stellar-origin black holes \citep{Abbott2021}. FSN is a possible explanation for these heavy black holes \citep[e.g.][]{Belczynski2020}.

Theoretically, the mechanism of explosion based on neutrino drives is still in the exploratory stage \citep{Connor2011,Ertl2016}. Besides, the evolution of massive star is sensitive to input physics, such as mass loss, metallicity, rotation, convection, binary interaction, and nuclear reaction rates \citep[see reviews by][]{Langer2012,2012ASSL..384..299H,Sylvia2021}, which can affect the evolution of internal chemical compositions and lead to different core structures at pre-collapse. Although there is currently no sufficiently precise criterion to determine whether the final product of evolution is a black hole or neutron star, it is generally believed that the core structure at pre-collapse is inextricably associated. \cite{Connor2011} proposed compactness parameter $\xi_M$, while \cite{Pejcha2015} used critical neutrino luminosity. In order to improve the accuracy of the criteria, a method of using two parameters is also proposed by \cite{Ertl2016}. Besides, \cite{Muller2016} proposed a semianalytic model describing the formation of a proto-neutron star, using five physical motivation parameters to determine the outcomes. This method not only provides the explodability of a given model but also yields other parameters of the explosion.

Theoretical calculations show that the explodability is very sensitive to the physical parameters of the model. \cite{Patton2020} showed the different outcomes for the same initial mass model by evolving carbon core with different initial CO core composition to iron core collapse. Convective burning events after the formation of CO core greatly affect the core structure at pre-collapse; the mass resolution and mesh quantity have also different degrees of impact on explodability \citep{Sukhbold2014}. Furthermore, mass loss as a crucial parameter of massive star evolution models, and has nonnegligible impact on the core structure. \cite{Renzo2017} systematically explored the sensitivity of massive stellar evolution to mass loss. They found that mass loss affects the core structure and the burning shell surrounding the core. During the main sequence, the core readjusts quasi-statically to accommodate mass loss, while subsequent evolutionary stages amplify the small variations generated during this phase.

The effect of rotation on the evolution of massive stars is complex and cannot be ignored. Stars with low metallicity are particularly sensitive to rotation because of a lower mass-loss rate. \cite{2000Heger} described a variety of instabilities caused by rotation that aid chemical mixing and angular momentum transfer, which means this constant rotation uniformizes the chemical composition of the star, also known as chemically homogeneous evolution \citep[CHE;][]{Maeder1987,Langer1992,Brott2011a}.
The most significant effect of rapid rotating is the increase in helium core mass, which is due to hydrogen being carried into the internal burning region. Multi dimensional simulations have shown that rotation induces a spiral arm structure during late burning phase, which may help with shock revive \citep{2014ApJ...785L..29M,2016ARNPS..66..341J,Chatzopoulos2016,Yoshida2021}.
Besides, observations have shown that a large number of massive stars have a fairly high rotation speed, and the highest one extends up
to $\sim$ 600 km $\rm s^{-1}$ \citep{Dufton2011,Agudelo2013}. It is necessary to include rotation when calculating explodability. However, rapid rotating stars may produce long gamma-ray bursts \citep[lGRBs;][]{Woosley1993,Yoon2006}, and it is not trivial to say that they look like FSN, since they produce luminous transients \citep{Kochanek2008}. Therefore, it is necessary to discuss the lGRBs scenario first.

In addition, about 7\% of OB stars have observable surface dipole magnetic fields with the order of kilogauss \citep{Fossati2015,Wade2016,Grunhut2017,Petit2019}. Although there is rare observational evidence \citep{2010ApJ...714L.318T,Oksala2012,2015MNRAS.451.2015O}, theoretical simulations suggest the interaction of the surface magnetic field with the stellar wind has a nonnegligible impact on the evolution of stars by limiting the wind material within closed magnetic field lines \citep{ud-Doula2002,Ud-Doula2004,Bard2016,Keszthelyi2019,Keszthelyi2020}, which leads to mass-loss quenching and magnetic braking, corresponding to reducing mass loss and angular momentum, respectively. 

In the present paper, we investigate the effects of rotaion, metallicity and magnetic field on FSN. In Section \S \ref{sec:Model} we briefly describe the input physics of the model.In Section \S \ref{sec:lGRBs}, we discuss models that might produce lGRBs. In Section \S \ref{sec:rotation}, we discuss the effects of rotation. In Section \S \ref{sec:Metallicity} we discuss the effects of magnetic fields. A brief summary is given in the Section \S \ref{sec:Conclusions}.

\section{Model}\label{sec:Model}
We use MESA stellar evolution code \citep[\textbf{version 10398}]{Paxton2011,Paxton2013,Paxton2015,Paxton2018,Paxton2019} to simulate stellar
evolution and structure from zero-age main sequence (ZAMS) to core collapse (the infall speed larger than 1000 km $\rm s^{-1}$ in the Fe core).
The present paper focuses on the effect of metallicity, rotation, and magnetic fields on the islands of explodability.
We select three different metallicities, $Z$ = 0.00034, 0.0017 and  0.0085, corresponding to 1/50 $ Z_\odot$, 1/10 $ Z_\odot$ and  1/2 $ Z_\odot$, respectively.
Here, $Z_\odot$ = 0.017, is the solar metallicity \citep{Grevesse1996}. Three different initial surface rotation velocities, 600, 300, and 0 km $\rm s^{-1}$, were selected to study the rotational effects. We also set up the models with magnetic fields that have a magnetic field strength of $10^3$ Gauss with an initial velocity of 600 km $\rm s^{-1}$.

In our models, the Ledoux criterion and the standard mixing length theory are used in convection simulation \citep{1958bohm},
where the mixing length parameter $\alpha_{\rm MLT}$= 1.5.
Overshooting is used for hydrogen burning stage with $\rm \alpha_{ov}$= 0.335, and the semiconvection mixing efficiency parameter $\rm \alpha_{SC}$ = 0.01 is applied.
The rotation can trigger the Goldreich--Schubert--Fricke instability, Eddington--Sweet circulation, dynamical instability, and secular instability, which produce the mixing \citep{Heger2000}.
Following \cite{Heger2000}, the ratio of the turbulent viscosity to the diffusion coefficient, $ f\rm_c$, is taken as 1/30.
We use \emph{approx21} nuclear network of MESA. The stellar winds used by \cite{Yoon2006,Marchant2016} are adopted by this work, in which the mass loss
from stellar wind can be calculated by the recipe in \cite{Vink2001} for hydrogen-rich stars (that is, the hydrogen abundance on the stellar surface $X_{\rm s}>$0.7),
while it can be done by the formula in \cite{Hamann1995} for hydrogen-poor stars ($X_{\rm s}<$0.4). The stellar wind of stars with $X_{\rm s}$ between 0.7 and 0.4 can be computed by the interpolation between the above two recipes. Following \cite{Bjorkman1993}, the mass loss enhanced by rotation is also included.
To avoid the rotation velocity exceeding the critical value, we limit $\rm \Omega/\Omega_{crit}$ to less than 0.98 \citep{Langer1998,David2020}.
\begin{figure*}
\includegraphics[width=1.0\textwidth]{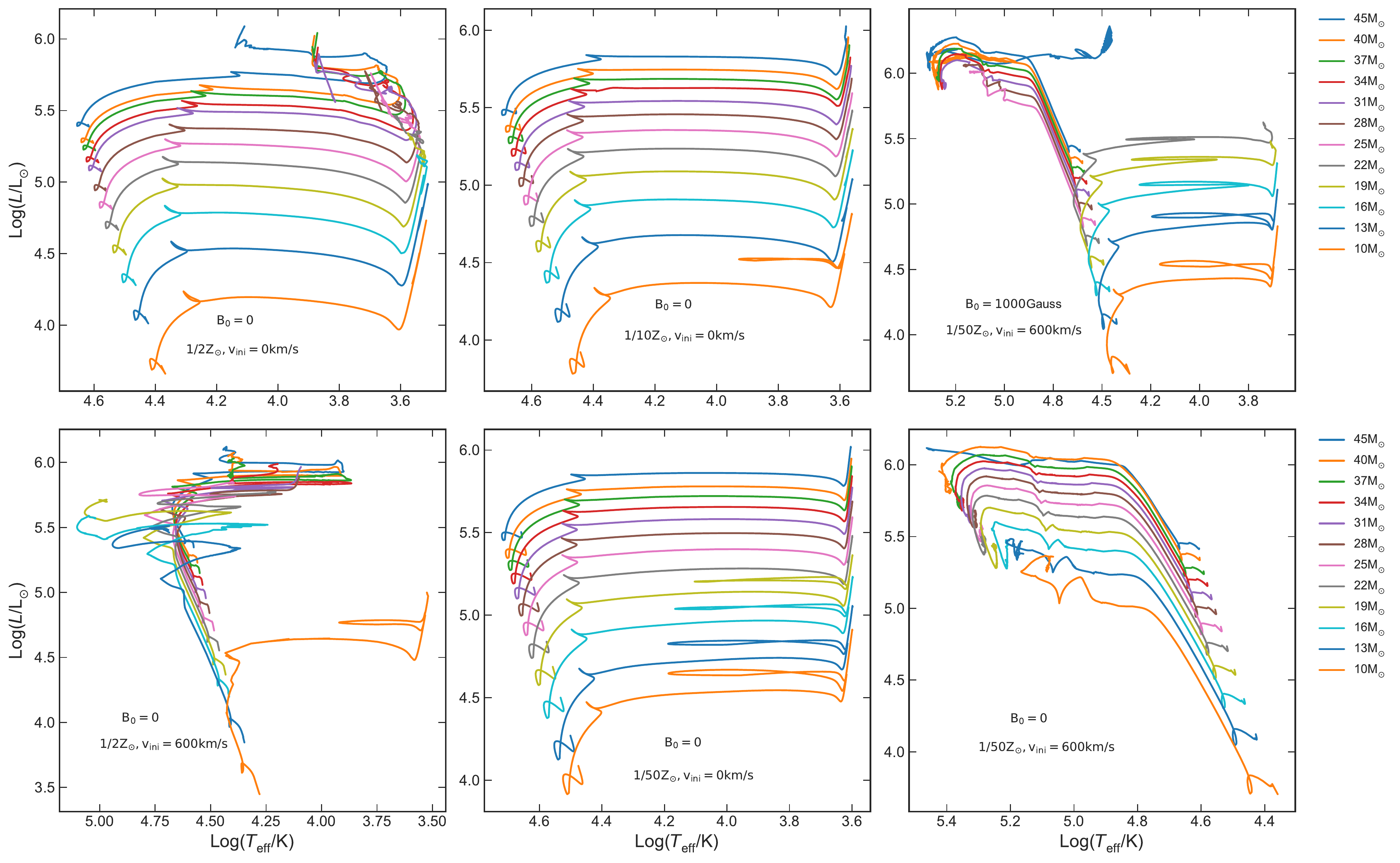}
\caption{The evolutionary tracks of the different model that contains various surface rotation velocity and metallicity, as well as surface fossil magnetic fields with 1000 Gauss.}
\label{fig:hr}
\end{figure*}

The origin and the effects of magnetic field on the stellar structure and evolution may be very complex.
For simplicity, we consider a surface fossil magnetic field.
Based on \cite{ud-Doula2009} and \cite{Keszthelyi2022}, the mass-loss quenching parameter $f_{\rm B}$ is described as
\begin{equation}
f_{\mathrm{B}}=\frac{\dot{M}}{\dot{M}_{B=0}}=1-\sqrt{1-\frac{1}{R_{\mathrm{c}}}} \quad \text { if } \quad R_{\mathrm{A}}<R_{\mathrm{K}}
\end{equation}
and
\begin{equation}
f_{\mathrm{B}}=\frac{\dot{M}}{\dot{M}_{B=0}}=2-\sqrt{1-\frac{1}{R_{\mathrm{c}}}}-\sqrt{1-\frac{0.5}{R_{\mathrm{K}}}} \quad \text { if } \quad R_{\mathrm{K}}<R_{\mathrm{A}}
\end{equation}
where $\dot{M}_{B=0}$ is the mass-loss rate, when $B=0$, $R_{\rm A}$ and $R_{\rm K}$ are the Alfv\'{e}n radius and the Kepler corotation
radius, respectively \citep[see][]{Petit2017}. $R_{\rm c}$ denotes the maximum distance of the last closed magnetic loop from the surface of the star and can be expressed as a function of the stellar radius $R_\ast$ and \emph{$R_A$} \citep{Petit2017,Keszthelyi2019,Keszthelyi2020}. It is approximated as $R_c \sim R_\ast + 0.7 ( R_A - R_\ast )$ \citep{ud-Doula2008,ud-Doula2009,Keszthelyi2022}.

The magnetic braking depletes the total angular momentum of star and slows down its rotation, and angular momentum decrease is calculated as
\begin{equation}
\frac{\mathrm{d} J_B}{\mathrm{~d} t}=\frac{2}{3} \dot{M}_{B=0} \Omega_{\star} R_{\mathrm{A}}^2,
\end{equation}
where $dJ_{\rm B}$/$dt$ and $\rm \Omega_{\star}$ are the rate of the system angular momentum loss and the surface angular velocity, respectively
\cite{Keszthelyi2020,Keszthelyi2022}.

\cite{Keszthelyi2020,Keszthelyi2022} proposed two schemes, the INT scheme and the SURF scheme, corresponding to calculating magnetic braking in the entire star and at the surface, respectively. The code used for the calculation is
\begin{equation}
\frac{\mathrm{d} J_{B}}{\mathrm{~d} t}=\int_{M_{t}}^{M_{\star}} \frac{\mathrm{d} j_{B}(m)}{\mathrm{d} t} \mathrm{~d} m=\sum_{k=1}^{x} \frac{\mathrm{d} j_{B, k}}{\mathrm{~d} t} \Delta m_{k},
\end{equation}\label{equ:m_b}
where $\mathrm \Delta m_{k}$ is the mass of a given layer, and $k$ is the index of layers inward of the model from the surface; the integration  from outside to inside determines the range used for calculations with index from $k$ = 1, to $k$ = x. For the INT case, the layers of the entire star are summed; however, we take the SURF case with $x$ = 500. Model retains most of the angular momentum to continue to evolve even after the main sequence with this case.

Figure \ref{fig:hr} shows H-R diagrams of stars with different input physics in our models.
The left two panels give the effects of the rotational velocity on the evolution of massive stars.
\cite{Brott2011a} presented a grid for the evolution of rotating massive stars. They found that rotation affects the evolutionary tracks of the star in the H-R diagram depending on the initial mass, rotation rate, metallicity, and the evolutionary stage. Since the mixing efficiency in our model is higher than that in \cite{Brott2011a},
the rotation is more efficient in the present paper. For nonrotating models with the same input parameters,
the evolutionary tracks in our models are similar to those in \cite{Brott2011a,2012A&A...537A.146E}.

The middle two panels of Figure \ref{fig:hr}  show the effects of metallicity. Comparing with 1/10  $Z_\odot$ models,
the effective temperature and luminosity of 1/50 $Z_\odot$ models are higher, which mainly results from the effects of the metallicity on the opacity; that is, the lower the metallicity is, the lower the opacity is.

The right two panels of Figure \ref{fig:hr} give the influence of magnetic field. In our models, a strong magnetic field restrains the mass-loss rate and slows down
stellar rotation. The stars with strong magnetic fields are systematically hotter and brighter, especially for the models that undergo ordinary evolution. This is consistent with the description of \cite{Keszthelyi2020}. The main reason is that the magnetic braking provides an additional mix to give the mean molecular weight lift and ultimately leads to an increase in luminosity. The increase in effective temperature is dominated by rotation-induced internal mixing \citep{Brott2011a}, even though the rotation is weakened by magnetic braking. In our simulations,
the 1/50 $Z_{\odot}$ models with mass lower than 22 $M_{\odot}$ have different evolutionary tracks from those with initial mass higher than 22 $M_{\odot}$.
The latter remains a thicker hydrogen-rich envelope at the end of the main sequence, while the former has a thinner  hydrogen-rich  envelope,
which results in them evolving naturally toward the bluewards of the H-R diagram and we discuss this in section \ref{sec:Metallicity}.

\section{Results}\label{sec:Results}
We simulate the evolution of massive stars that include rotation and magnetic field with initial mass of 10 $M_{\odot}$ to 55 $M_{\odot}$ and the interval is 1 $M_{\odot}$. Figure \ref{fig:hr} shows the evolutionary tracks of some models in the H-R diagram, and below, we discuss the explodability at core collapse.

\subsection{Explodability Criterion}\label{sec:Explodability}

It is very difficult to judge the explodability of massive stars. \cite{Connor2011} introduced a compactness parameter, $\xi_{M}$, which is  mainly used to quantify density gradients within a mass coordinate and defined by
\begin{equation}\label{compactness}
\xi_{\rm M}=\frac{M/\rm M_{\odot}}{R(M_{\rm bary}=M)/1000\, {\rm km}},
\end{equation}
and $M$ = 2.5 $M_{\odot}$ is used to quantify density gradients near the iron core. The compactness parameter is generally calculated when the collapse speed reaches 1000 km $\rm s^{-1}$. Although $\xi_{2.5}$ is originally intended to predict the final fate of a nonrotation model, it is still available for fast rotators \citep[e.g.][]{David2020}. The rotational energy provided by rotation may contribute to the explosion of the model, which reduces the accuracy of compactness, but it still provides enough valid information about the internal structure of the model at core collapse. $\xi_{2.5}$ = 0.45 is used to distinguish between explosion or implosion.
In this work, following \cite{David2020},  we assume that a successful explosion occurs when $\xi_{2.5} < 0.45$; otherwise,an FSN occurs.

However, the explosion of massive stars is complex, and the core structure of a single parameter alone is limited.
\cite{Ertl2016} proposed a relationship between the mass of the core and the density gradient of stars at core collapse. They define two variables, $M_4$ and $\mu_4$, and use them to parameterize the core structure of the star. They are defined as $M_4$ = m ($s$ = 4 $\rm k_B$), the mass coordinate where specific entropy reaches 4 $\rm k_B$, and the radial density gradient
\begin{equation}
\mu_{\rm 4}=\frac{{\rm d}m/\rm M_{\odot}}{{\rm d}r/1000\, {\rm km}},
\end{equation}
$M_4$ is chosen arbitrarily but is found to be a good proxy of the location of the oxygen shell \citep{Sukhbold2018}, and $\mu_4$ is evaluated by taking d$m$ = 0.3 $M_{\odot}$, and is a proxy of the mass gradient at the edge of the core.
In this work, we discuss the effects of $\xi_{2.5}$, $\mu_4$  and $M_4$ on judging the explodability.

 \subsection{Explodability and Long Gamma-Ray Bursts}\label{sec:lGRBs}
\begin{figure*}[ht]
  \subfigure{\includegraphics[width=1.0\textwidth]{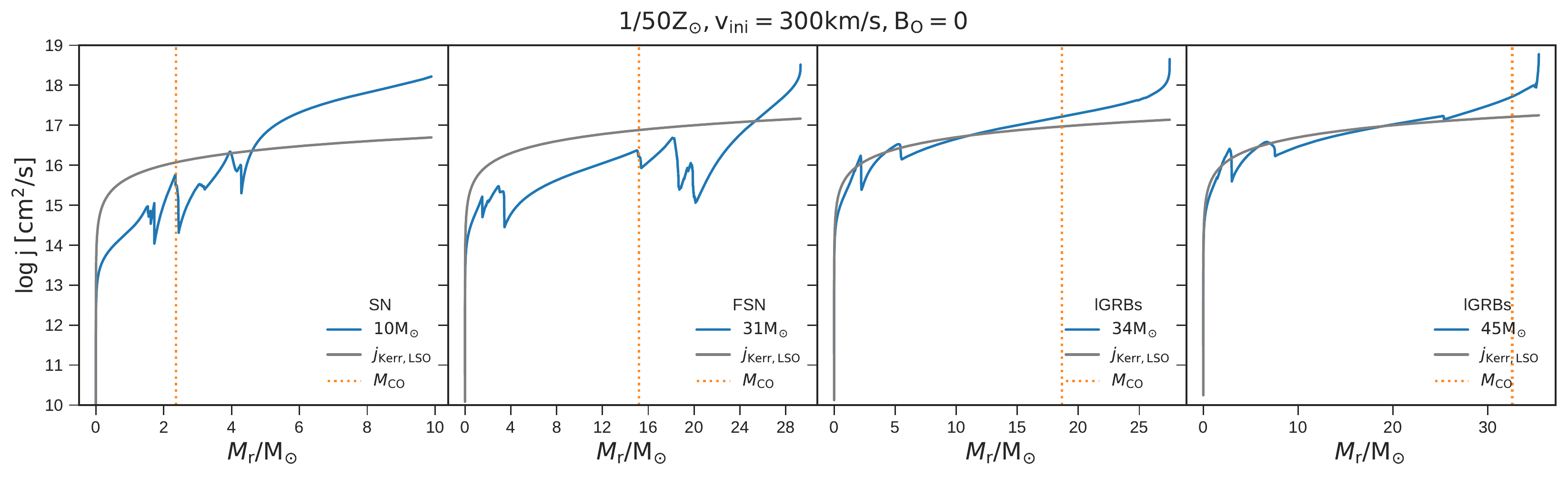}}
  \caption{Internal specific angular momentum distribution of 1/50 $Z_{\odot}$ models with 300 km s$^{-1}$ and without magnetic fields at core collapse. The blue, gray, and dotted lines represent the specific angular momentum, $j_{\rm LSO}$ and CO core masses of the model, respectively.}
  \label{fig:lGRB}
\end{figure*}

The present paper investigates the explodability of rotating massive stars.
It is well known that rapidly rotating massive stars that undergo CCSNe can produce lGRBs \citep{Woosley1993,Yoon2006}, which yield bright transients that do not match the observational definition of FSN.
According to \cite{Yoon2006}, lGRBs are generated when specific angular momentum ($j_{\rm co}$) at any location in the CO core is greater than the
last stable orbit ($j_{\rm Kerr, LSO}$) \citep{Bardeen1972} at core collapse.
We adopt this criterion to explore the regions of the model where lGRBs occur and exclude them from FSN.
Therefore, in the present paper, the conditions for an FSN occurring are as follows: $\xi_{\rm 2.5}>0.45$ and $j_{\rm co}<j_{\rm Kerr, LSO}$ at core collapse. And the conditions for an lGRB occurring are $\xi_{\rm 2.5}>0.45$ and $j_{\rm co}\ge j_{\rm Kerr, LSO}$.

Figure \ref{fig:lGRB} shows the different  compact remnants (successful SN, FSN, lGRB) for 1/50 $Z_{\odot}$ models with initial rotation velocity of 300 km s$^{-1}$. The lower mass models with rotation velocity of 300 km s$^{-1}$ do not necessarily undergo CHE (see Figure \ref{fig:hr}), the expansion of the envelope further slowing down the rotation and resulting in internal specific angular momentum of CO core lower than $j_{\rm LSO}$ at core collapses as shown in Figure \ref{fig:lGRB}.
Since a star needs to experience CHE to retain enough specific angular momentum to produce a lGRB \citep{Yoon2012}, lGRBs are produced at $M_{\rm ZAMS}$ $\ge$ 34 $M_{\odot}$, which corresponds to CHE model of Figure \ref{fig:hr}.
The models with a rotation velocity of 600 km s$^{-1}$ usually leave a bare CO core at core collapse, and retain enough specific angular momentum to form lGRBs when the black hole formed \citep{David2018,David2020}. Besides,
none of the 1/10 $Z_{\odot}$ models will produce lGRBs due to the increased mass loss, and more angular momentum is carried away.
The models that include surface magnetic fields show similar behavior.  
\subsection{Explodability of Rapidly Rotating Massive Stars }\label{sec:rotation}
\begin{figure*}
  \centering
  \includegraphics[width=1.0\textwidth]{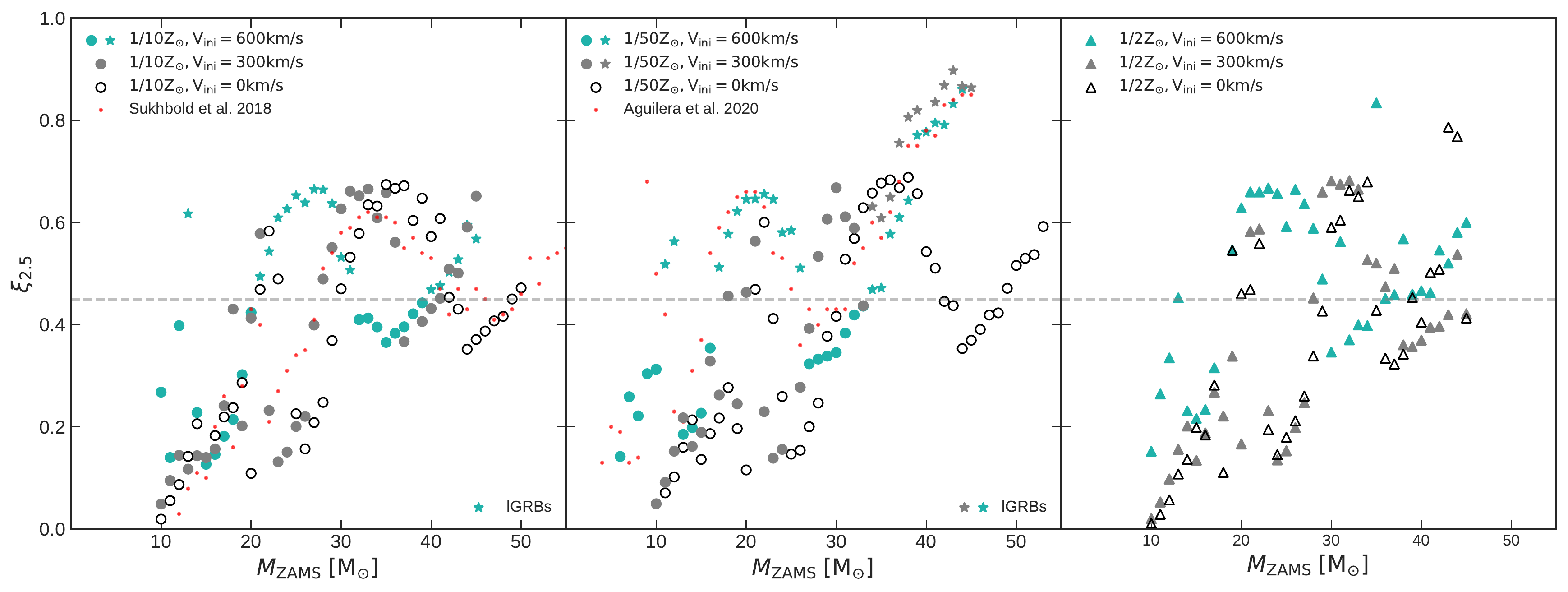}\\
  \caption{Compactness parameter $\xi_{2.5}$ as a function of initial mass at core collapse of different initial models. The dotted line at $\xi_{2.5}$ = 0.45 separates the models that might explode (below the line) or implode (above the line) according to \cite{Connor2011}. LGRBs (star markers) are generated in implosion models where $j_{\rm co}\ge j_{\rm Kerr, LSO}$. Taking into account the change curve of the compactness parameter, the model with 1/50 $Z_{\odot}$ and $V_{\rm ini}$ = 600 km $\rm s^{-1}$ is calculated from 6 $M_{\odot}$. The maximum initial mass of the nonrotation model is 55 $M_{\odot}$, and the other models range from 10 to 45 $M_{\odot}$. The mass intervals of all models are 1 $M_{\odot}$. The results for the nonrotating models calculated by \cite{Sukhbold2018} are given by the red dots in the left panels, while those for
  the rapidly rotating models ($V_{\rm ini}=600$ km s$^{-1}$) calculated by \cite{David2020} are shown by the red dots in the middle panels. CHE models are represented by pentagram markers, including all rapid rotating models and 300 km $\rm s^{-1}$ models with $M_{\rm ZAMS}\ge$ 34 $M_{\odot}$.
  }\label{fig:kexi}
\end{figure*}

As mentioned in the above section, the explodability in this work is determined by $\xi_{\rm 2.5}$.
Figure \ref{fig:kexi} shows $\xi_{\rm 2.5}$ as a function of initial mass at core collapse for the models with
different metallicities and initial rotational velocities.

Based on the compactness parameter, $\xi_{\rm 2.5}$,  \cite{Sukhbold2018} investigated the explodability of nonrotating models with solar metallicity, and they found that massive stars with a mass between about 20 $M_{\odot}$ and 22 $M_{\odot}$ or larger than about 28 $M_{\odot}$ would fail to explode.
The similar islands have been obtained by other groups \citep{Ugliano2012,Ivanov2021}.
In our work, the islands of FSN for the nonrotating models appear between about 21 and 23 $M_{\odot}$ or above about 30 $M_{\odot}$, which is almost
consistent with the results in  \cite{Sukhbold2018}.

In order to investigate the superluminous SNe and lGRB, \cite{David2020} calculated $\xi_{\rm 2.5}$ for rapidly rotating massive stars with $Z$ = 1/50 ${Z}_\odot$,
which are given by the red dots in the middle panel of Figure \ref{fig:kexi}. Our results of the rapidly rotating models are similar to theirs.
Compared with the nonrotating models, the rapidly rotating massive stars have larger $\xi_{\rm 2.5}$.
It means that the island of  explodability should enlarge. However, due to $j_{\rm co}> j_{\rm Kerr, LSO}$,
in our models with $V_{\rm ini}=600$ km s$^{-1}$, these massive stars do not undergo FSN but become lGRB.

Intriguingly, in our models with $V_{\rm ini}=300$ km s$^{-1}$,  all massive stars with $Z$ = 1/10 ${ Z}_\odot$ cannot evolve into lGRB, while these with mass higher than about 34 $M_{\odot}$ and $Z$ = 1/50 ${Z}_\odot$
become lGRBs.
The main reason is that the rotation in a low-metallicity model can trigger an efficient CHE.
\subsubsection{Effects of metallicity}
Metallicity has an effect on rotational velocities, surface abundance, advanced evolutionary stages, and evolutionary tracks in the HR diagram \citep{Brott2011a,Chieffi2013,Choi2016,Yoon2017,Groh2019}. As discussed in Section \ref{sec:Model}, metallicity affects the minimum mass of models experience CHE, which changes the distribution of compact remnants.
For instance, efficient CHE occurs more difficultly  as metallicity increases. For the initial velocity of 300 km $\rm s^{-1}$, the models with higher mass ($M_{\rm ZAMS}\ge$ 34 $M_{\odot}$) of 1/50 $Z_{\odot}$ can efficiently undergo CHE; however,
all of 1/10 $Z_{\odot}$ models undergo normal evolution. This mainly results from two factors. Firstly, the enhanced mass loss causes more angular momentum to be carried away. Secondly, the stars with higher metallicity become less compact, which results in a longer mixing time-scale \citep{Keszthelyi2022}.

As discussed in Section \ref{sec:lGRBs}, in the present paper, a star that experiences efficient CHE will always end up as collapsar and form lGRB, with the exception of 1/2 $Z_{\odot}$. As the right panel of Figure \ref{fig:kexi} show, for 1/2 $Z_{\odot}$ models with velocity of 600 km $\rm s^{-1}$,
none of them becomes lGRBs because high metallicity greatly suppresses the mixing effect \citep{Yoon2012}.

\subsubsection{Effects of rotation}
\cite{Maeder1987} found that, if rotational mixing occurs at a faster rate than the formation of chemical gradients driven by nuclear fusion in massive main-sequence stars, it can lead to completely different evolutionary behavior in the stars throughout the core hydrogen burning phase. All compositional gradients are smoothed out, and the composition of the star becomes essentially homogeneous everywhere. This enables the star to smoothly transition into a helium star of the same mass. Having retained little or no hydrogen envelope at formation, this helium star, rather than expanding after main sequence as in normal stellar evolution, it continues contracting in radius and evolves blueward in the H-R diagram \citep{2005A&A...443..643Y,David2018}.

\begin{figure*}
  \includegraphics[width=1.0\textwidth]{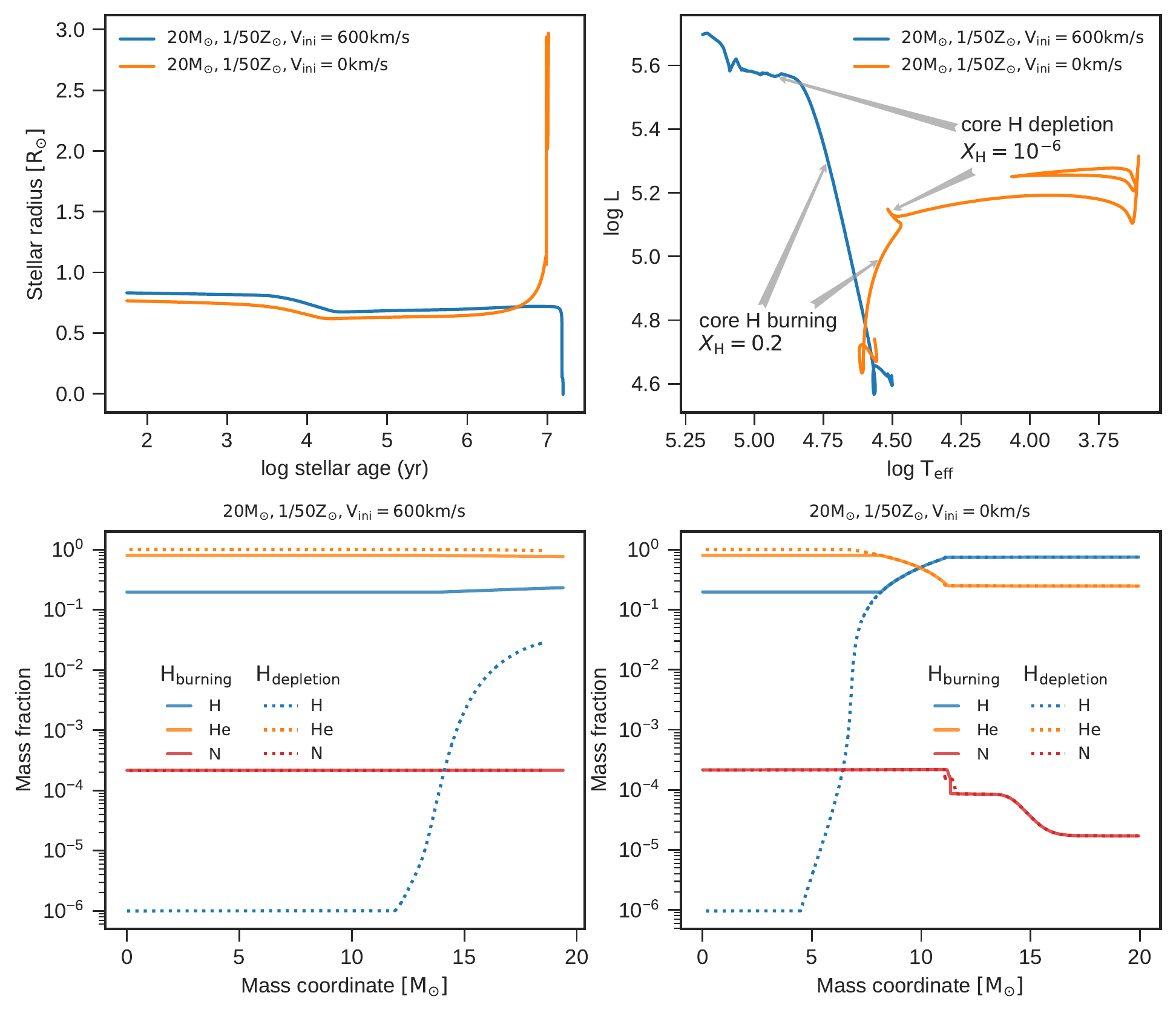}\\
  \caption{Verifying chemically homogeneous evolution of low-$Z$, rapid rotators. We use 20 $M_\odot$ models with and without rotation as examples, with identical inputs other than initial rotational velocity. We evolve these two models from ZAMS to central H exhaustion. The upper left panel shows the evolution of stellar radius with time. The lower left is the HR diagram. The two middle panels respectively show the internal abundance distributions of the rapidly rotating model and nonrotating model during the main sequence, while the two right panels show the internal abundance distributions when central H is exhausted. 
  }\label{fig:che}
\end{figure*}

Figure \ref{fig:che} shows our rapidly rotating model as having undergone CHE.
The upper left and right panels show that the rapid-rotating model (in orange) evolves blueward in the H-R diagram and continues contracting after terminating the main sequence, instead of expanding in radius and evolving redward like the nonrotating model (in blue) after the  main sequence. 
In addition, we have annotated the locations at the stage of core hydrogen burning (in this case, we select the point at which the core H mass fraction reaches 0.2) and exhaustion with arrows in the H-R diagram. The lower left and right panels separately illustrate the element abundance distributions within these two models at these two evolutionary stages, depicting their internal chemical evolution during the main sequence. 
As shown in the lower left panel, the rapidly rotating model exhibits nearly homogeneous chemical composition during the  main sequence evolutionary stage. Upon transitioning into helium stars, the surfaces of this model are nearly devoid of hydrogen. This stands in stark contrast with the two rightmost panels showcasing nonrotating models (lower right panel). In nonrotating model, distinct stratification of chemicals exists within the stars during the main sequence, and appreciable hydrogen remains on their surfaces as He stars emerge. 
Figure \ref{fig:che} illustrates that these characteristics are consistent with those discussed previously for the CHE model.

\begin{figure*}
  \centering
  \includegraphics[totalheight=5.2in,width=5.5in]{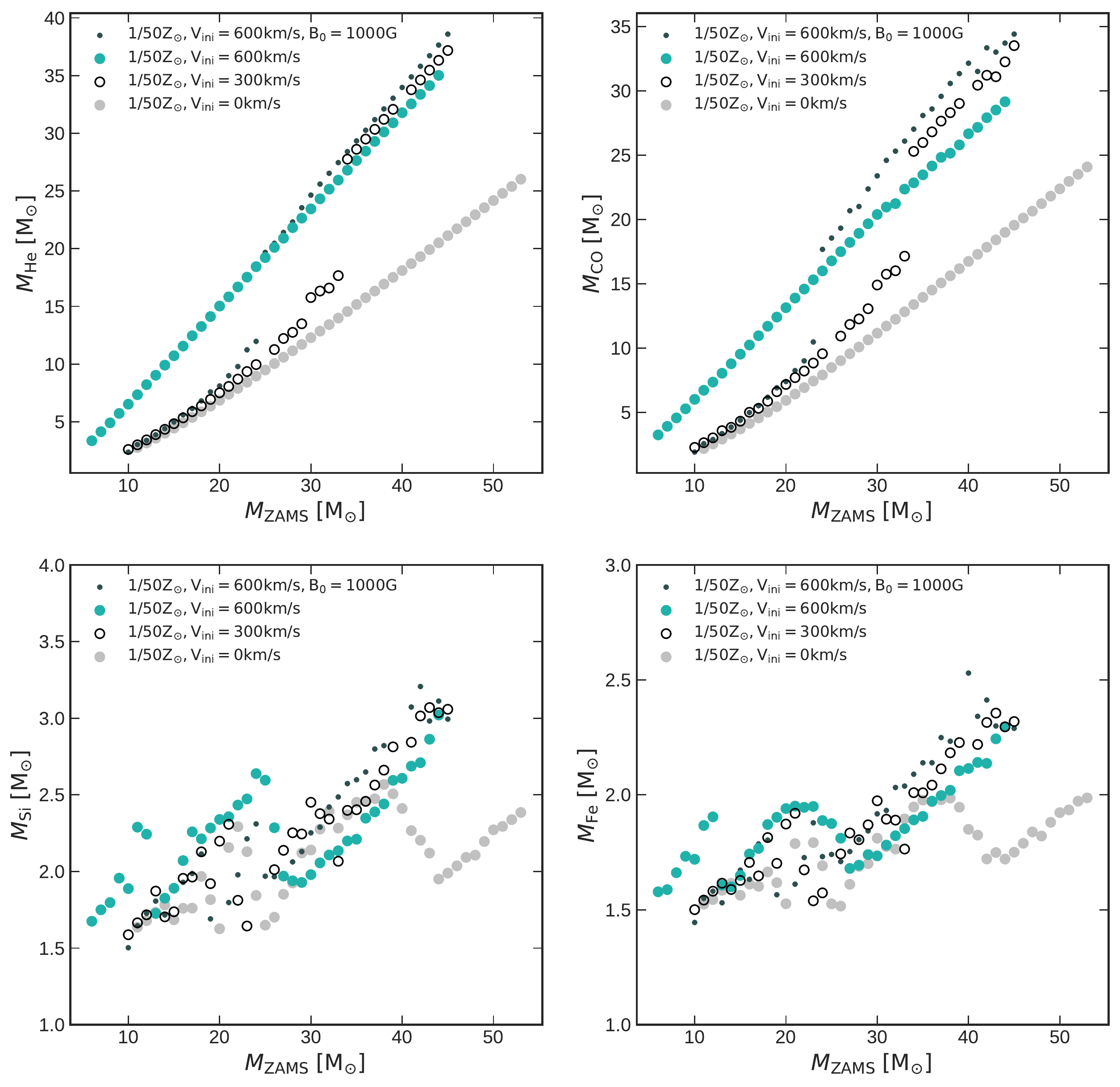}
  \caption{He core, CO core, Si core, and Fe core as function of initial mass for the different initial rotational velocities at the core collapse.
  Only sets with metallicity of 1/50 $Z_{\odot}$ are selected to explore the effects of rotation. Different initial surface rotations are represented by different markers. 
  }\label{fig:core}
\end{figure*}

In order to discuss the effects of rotation on the explodability, we take the models with $Z=1/50 {Z_{\odot}}$ as examples.
Figure \ref{fig:core} shows the masses of He, CO, Si, and Fe cores for the different initial rotational velocities at the core collapse.
Due to the CHE triggered by rapid rotation, the models with $V_{\rm init}=600$ km s$^{-1}$ have larger He and CO cores than those without rotation.
For the models with $V_{\rm init}=300$ km s$^{-1}$, the stars with $M_{\rm ZAMS}$ larger than about 34 $M_\odot$ have undergone efficient CHE, and have larger He and CO cores than those with  $M_{\rm ZAMS}$ smaller than about 34 $M_\odot$. The main reason is as follows:
the larger the stellar mass is, the larger its radius is. Then, its Keplerian velocity is smaller. Therefore, the CHE is more efficient in models with larger masses for the same initial rotational velocity. 
Moreover, the transition between normal evolution and CHE affects the properties of compact remnants differently. 
For instance, magnetic models with initial masses greater than 25 $M_{\odot}$ undergo efficient CHE, resulting in a growing He core during the main sequence that subsequently shrinks before central He ignition due to the absence of a hydrogen envelope. In contrast, the 22 $M_{\odot}$ model exhibits weak rotation-induced mixing, and its central convective zone gradually fades after the start of the main sequence due to the decrease in central H abundance. Consequently, the He core of the 22 $M_{\odot}$ model is substantially smaller than that of the 25 $M_{\odot}$ model. 

Obviously, both He and CO cores behave monotonously with the initial mass, while Si and Fe cores are nonmonotonic in behavior.
The latter is similar with that of $\xi_{2.5}$ shown by Figure \ref{fig:kexi}.
This means that the effect of rotation on $\xi_{2.5}$ can be projected to changes in the mass of the iron core \cite[also see][]{Sukhbold2014,David2020};
that is, rotation causes the helium core to enlarge, and the larger helium core affects the advanced burning phases,
eventually forming heavier iron core and higher compactness parameter.
 
\begin{figure*}
    \centering
    \includegraphics[totalheight=2.8in,width=5.5in]{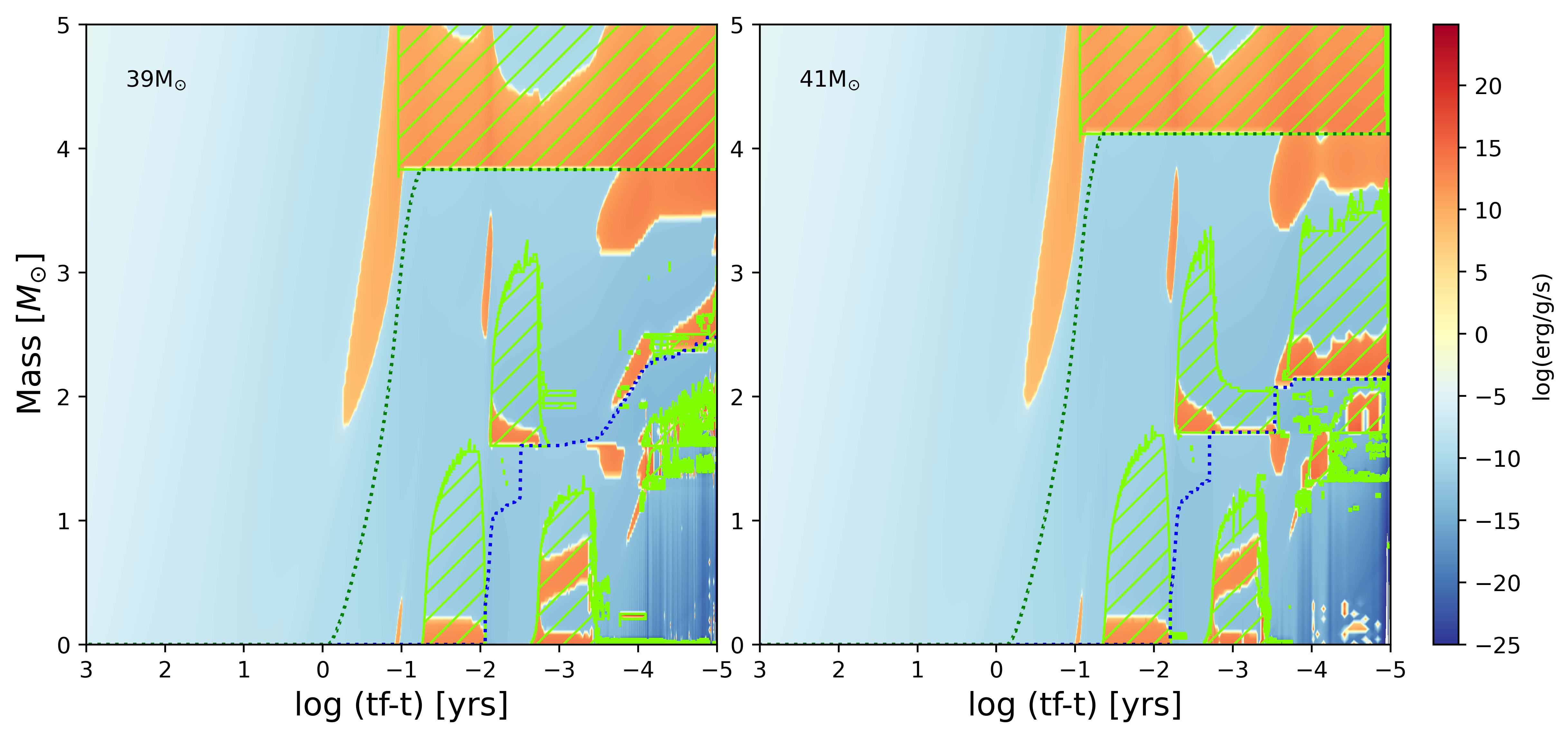}
    \caption{Convection histories of two nonrotating models with 1/50 $Z_{\odot}$, representing the structure of convective regions changes that are responsible for the significant variations of compactness. The structure of the innermost 5 $M_{\odot}$ material as a funciton of time remaining until core collapse, from core carbon burning until core collapse. Colors denote energy generation (red) and energy loss (blue) gradients. Green diagonal area denote convective regions. Green dot line and blue dot line indicate the mass of O core and Si core, respectively. Closely related to the Si core mass are the three oxygen burning episodes (including a central core-burning episode) above the blue line. Their $M_{\rm ZAMS}$ are close; however, 39 $M_{\odot}$ will experience FSN while 41 $M_{\odot}$ successfully explodes.
    }
	\label{fig:convection}
\end{figure*}

Notably, the explosibility is highly sensitive to the advanced evolutionary stages of the model. Figure \ref{fig:convection} shows the convective histories of two low-metallicity, rapidly rotating models after central helium depletion. Despite the former having a lower $M_{\rm ZAMS}$, its $\xi_{2.5}$ exceeded 0.45, and it will fail to explode; the latter had a higher $M_{\rm ZAMS}$ but $\xi_{2.5} <$ 0.45, enabling a successful explosion. The primary reason is that, for the latter, hours before core collapse, there remained a sizable convective region outside the Si core (third green-shaded region on the blue line in the right panel), keeping the mass of the Si core always below 2.5 $M_\odot$ (see blue line in right panel). 
While the former is only a very weak convective region outside the Si core (left subgraph blue line), which makes the mass of Si core increase by more than 2.5 $M_{\odot}$ (see left subgraph blue line). Therefore, $\xi_{2.5}$ is closely related to the convective region history of massive stars, which is consistent with \cite{Sukhbold2018}.

In addition, rapid rotation makes the transition, from convective carbon burning to radiative burning, occur in the models with lower initial mass.
For the non-rotaion model, the transition usually occurs around at $\sim$ 21 $M_{\odot}$ \citep{Sukhbold2018}.  However, as Figure \ref{fig:kexi} has shown, the transition in the models with rapidly rotation occurs at $\sim$ 11 $M_{\odot}$.

\begin{figure}
  \includegraphics[width=\columnwidth]{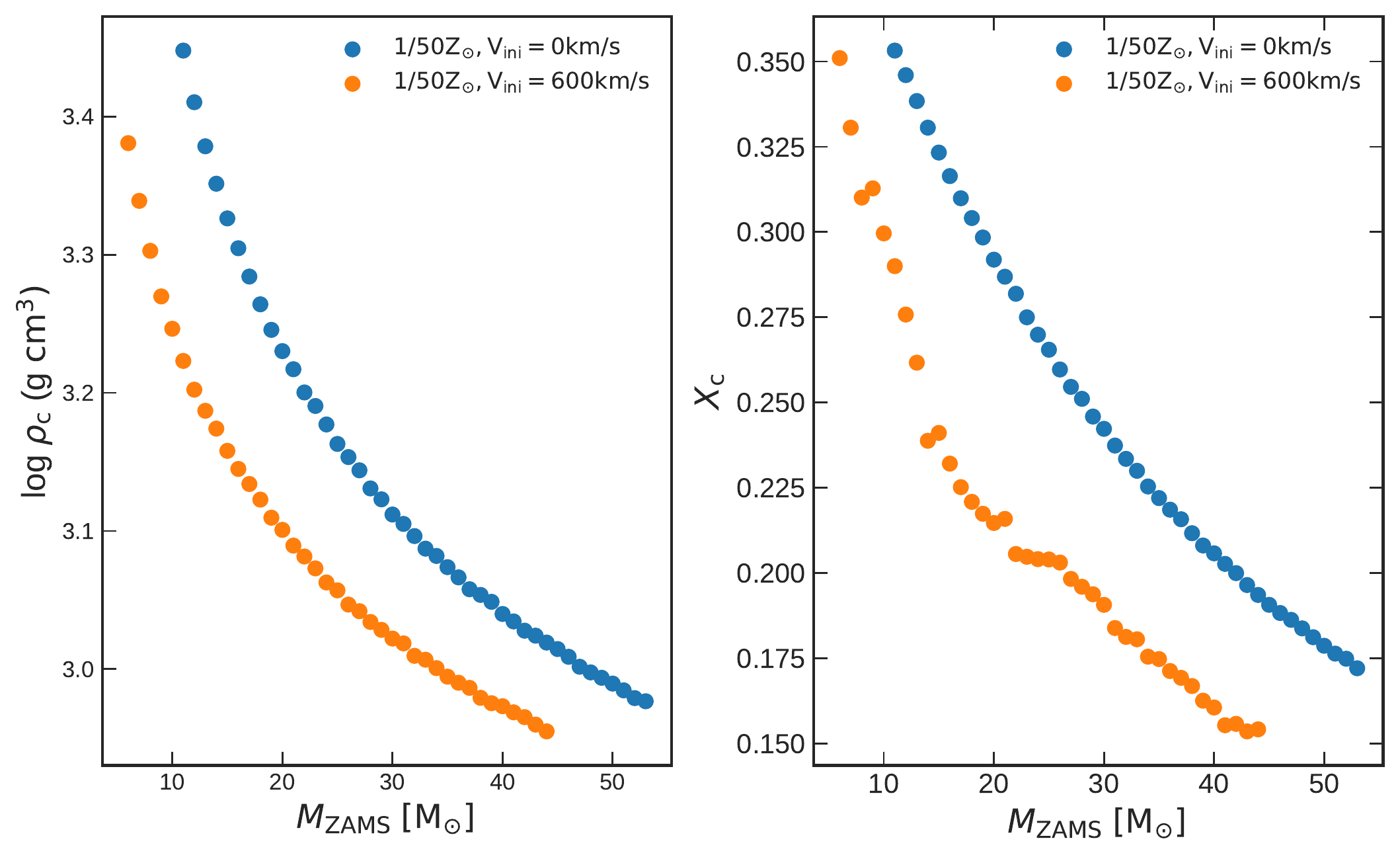}
  \caption{Central density and $\rm ^{12}C$ mass fraction at the time of central helium depletion. }\label{fig:c12}
\end{figure}

This can be explained by the larger He core of the rapid rotating models.
Naturally, the subsequent helium burning products of rotating models distinguish from the products of nonrotating models  because of the competitive  reactions between
$\rm 3\alpha$ and $\rm ^{12}C(\alpha,\gamma)^{16}O$ \citep{Tur2007,Ekstrom2010}.
Figure \ref{fig:c12} gives the central density and faction of central $\rm ^{12}C$ at the end of central helium depletion. Obviously, the fraction of $\rm ^{12}C$ in the rapid rotating models is much lower than that in the non-rotaing models.

According to \cite{deBoer2017}, the molar abundance $Y (\rm ^{12}C)$ can be calculated as a function of helium seed abundance $Y (\rm ^{4}He)$, helium burning process reaction rate $\lambda$, and density $\rho$ using the equation
\begin{equation}
\begin{aligned}
\frac{\mathrm{d} Y\left({ }^{12} \mathrm{C}\right)}{\mathrm{d} t}=& \frac{1}{3 !} Y^{3}\left({ }^{4} \mathrm{He}\right) \cdot \rho^{2} \cdot \lambda_{(3 \alpha)} \\
&-Y\left({ }^{4} \mathrm{He}\right) \cdot Y\left({ }^{12} \mathrm{C}\right) \cdot \rho \cdot \lambda_{^{12} \mathrm{C}(\alpha, \gamma){ }^{16} \mathrm{O}} \text{.}
\label{equ:C}
\end{aligned}
\end{equation}
Equation \ref{equ:C} shows the key factors of $\rm ^{12}C$ mass fraction at central helium depletion, which is central density and reaction rate of 3$\alpha$ and $\rm ^{12}C(\alpha,\gamma)^{16}O$.
The rapid rotating models remain lower in central density and higher central temperature during the central helium burning phase \textbf{(Figure \ref{fig:c12} left panel)}, which results in the consumption of $\rm ^{12}C$  being faster by $\rm ^{12}C(\alpha,\gamma)^{16}O$ than producing it by 3$\alpha$ and finally getting a lower $\rm ^{12}C$ mass fraction at central helium depletion.
\cite{Sukhbold2020} suggested that the critical condition of the transition is related to the central $^{12}{\rm C}$ mass fraction at the time of carbon ignition. According to Figure \ref{fig:c12}, the transition of rapid rotating models will occur at lower $M_{\rm ZAMS}$ since the central $^{12}{\rm C}$ mass fraction is lower than that of nonrotating models.

\begin{figure}
  \includegraphics[width=\columnwidth]{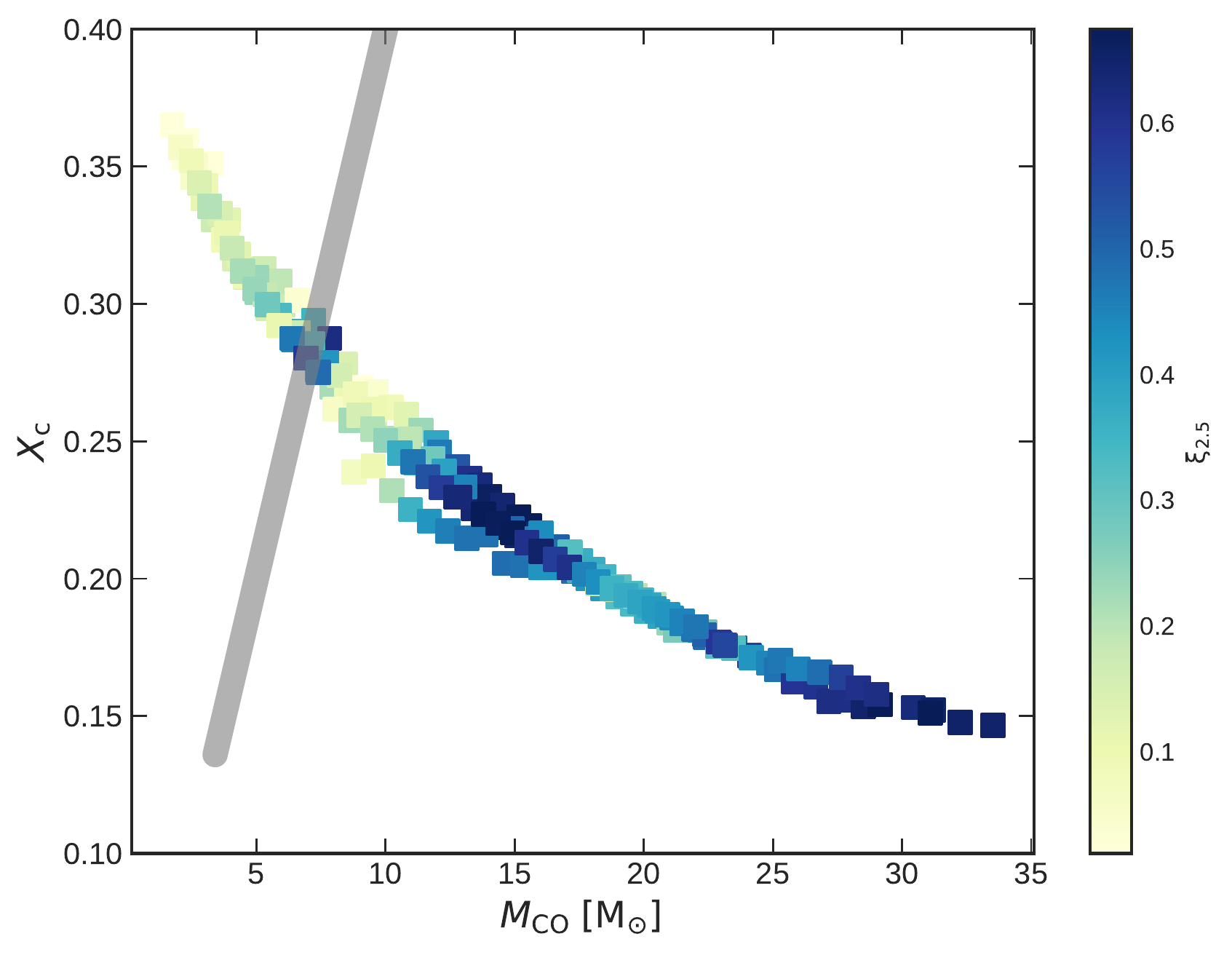}
  \caption{CO core structures superimposed on the $\xi_{2.5}$ distribution for the models with different metallicities and rotations when central helium depletes.
  The gray shaded area roughly corresponds to the transition of the C burning mode from convection (left) to radiation (right) according to \cite{Patton2020}.
  The bar is truncated at a CO core mass of $3.2M_\odot$ and a central carbon mass fraction of $0.14$ because, below these values, carbon burning proceeds in the convective region in their work. 
  }\label{fig:s20}
\end{figure}
\cite{Patton2020} constructed a grid of CO core models tracking the evolution from carbon ignition to core collapse. Their Figure 2 illustrates a loose boundary demarcating the transition from convective to radiative carbon burning. 
Our Figure \ref{fig:s20} exhibits $\xi_{2.5}$ for models of varying metallicities and initial rotational velocities; larger values (darker colors) indicate greater difficulty in explosion. The gray shaded region denotes the boundary as specified in \cite{Patton2020}. Intriguingly, the carbon burning transition boundary for our models, both with and without rotation, agrees closely with theirs. This suggests that the transition of carbon burning mode depends almost entirely on the CO core structure at central carbon ignition, and the rotation appears to have little effect. For example, when the metallicity is 1/50 $Z_\odot$, the transition for a nonrotating model occurs at a CO core mass of 6.92 $M_\odot$ and central $^{12}$C mass fraction of 0.28, while for a rapidly rotating model it takes place at a CO core mass of 6.73 $M_\odot$ and central $^{12}$C mass fraction of 0.29.

\cite{Limongi2018} presented a grid of massive star models with different metallicity and
rotation speeds included. Due to a wide initial-mass interval, they gave a smoothly monotonic relation between the initial mass
and $\xi_{2.5}$. However, \cite{Chieffi2020} showed a finer grid of massive star models, and found a nonmonotonic trend between
the initial mass and $\xi_{2.5}$. They also compared this trend with that of Sukhbold \cite{Sukhbold2018}, and considered that the large difference
originates from the various convective episodes. As shown by Figure 3, our results are similar with the trend of \cite{Sukhbold2018}.

\subsubsection{The Effects of the other input parameters}
Considering that small changes in the input parameters can lead to dramatic changes in compactness, we explore the effects of some input parameters including varying the mass and time resolution, and the convective boundary mixing. In MESA, the mass and time resolution are controlled by two input parameters \textsf{mesh\_delta\_coeff} and \textsf{varcontrol\_target}, respectively. The convective boundary mainly is determined by 
the convective overshooting parameter ($\alpha_{\rm OV}$) and the mixing length parameter ($\alpha_{\rm MLT}$).
Table \ref{tab} shows the evolutional results of fast-rotating models with an initial mass of 20 $M_\odot$ and metallicity 1/50 $Z_\odot$, 
undervarying the above input parameters.
It gives the zone number and the step number for simulating stellar structure and evolution, 
the stellar mass and radius, the masses of He core, Fe core, CO core, and the compactness at pre-collapse, and the mass $^{12}{\rm C}$ at the central C ignition.

The first three models (\texttt{20a}-\texttt{20c}) represent models with different mass and time resolutions. Model \texttt{20a} represents the setup adopted in this paper, while the models \texttt{20b} and \texttt{20c} correspond to calculations using larger time steps and less refined grids, respectively. They combine zoning and time-resolution choices to examine the impact on compactness and other properties. The core structure before core collapse does not differ significantly between these models. 
Compared to the impact of rotation, the effects of mass and time resolution appear negligible.

Table \ref{tab} also illustrates the model with convective overshooting turned off. 
Although the helium core mass in this model is larger before the collapse, its compactness does not vary much.
In fact, during the main sequence, the overshooting increases the helium core mass. Therefore, the models without convective overshooting have a smaller helium core mass at central hydrogen exhaustion and retain a higher surface hydrogen abundance. The mass-loss scheme of \cite{Hamann1995} depends on the surface hydrogen abundance, and a higher surface hydrogen abundance means lower mass loss. Therefore, before the collapse, the models without convective overshooting retain slightly higher masses. Simultaneously, we also explore models adopting even larger values of $\alpha_{\rm MLT}$, which may slightly influence the convection boundary but will not cause significant changes to the results either.

In short, while selecting different input parameters (such as mass (\textsf{mesh\_delta\_coeff}), time resolution (\textsf{varcontrol\_target}), and overshooting) in the model can affect the core structure at pre-collapse very weakly, compared with the effect of rotation, these differences are negligible.

\begin{deluxetable*}{cccccccccccccc}
  \tablecaption{Presupernova properties with varying resolution for the 20 $M_{\odot}$, 1/50 $Z_\odot$ rapidly rotating model.}
  \tablenum{1}
  \tablehead{\colhead{$M_{\rm ZAMS}$} & \colhead{$\delta_{\rm mesh}$} & \colhead{varcontrol\_target} & \colhead{$\alpha_{\rm OV}$} & \colhead{$\alpha_{\rm MLT}$} & \colhead{$N_{\rm zones}$} & \colhead{$N_{\rm steps}$} & \colhead{$M_{\rm preSN}$} & \colhead{$R_{\rm preSN}$} & \colhead{$M_{\rm He}$} & \colhead{$M_{\rm CO}$} & \colhead{$\rm ^{12}C_{ign.}$} & \colhead{$M_{\rm Fe}$  } & \colhead{$\xi_{2.5}$} \\ 
\colhead{($M_{\odot}$)} & \colhead{} & \colhead{}& \colhead{} & \colhead{} & \colhead{} & \colhead{} & \colhead{($M_{\odot}$)} & \colhead{($R_{\odot}$)} & \colhead{($M_{\odot}$)} & \colhead{($M_{\odot}$)} & \colhead{} & \colhead{($M_{\odot}$)} & \colhead{} }
\startdata
20a &0.5& 3$\times$10$^{-4}$& 0.335 & 1.5 & 3598 & 2613 & 15.64 & 0.45 & 15.64 & 13.22 & 0.19 & 1.93 & 0.63 \\
20b &0.5 &3$\times$10$^{-3}$& 0.335 & 1.5 & 3612 & 2287 & 15.64 & 0.44 & 15.64 & 13.21 & 0.19 & 1.90 & 0.64 \\
20c &1.5& 3$\times$10$^{-4}$& 0.335 & 1.5 & 2088 & 3100 & 15.61 & 0.43 & 15.61 & 13.21 & 0.19 & 1.88 & 0.58 \\
20d &0.5& 3$\times$10$^{-4}$& 0 & 1.5 & 3703 & 3336 & 15.84 & 0.48 & 15.84 & 13.28 & 0.20 & 1.94 & 0.64 \\ 
20e &0.5& 3$\times$10$^{-4}$& 0.335 & 2.0 & 3605 & 2356 & 15.64 & 0.44 & 15.64 & 13.21 & 0.19 & 1.90 & 0.64\\
\enddata
\tablecomments{Presupernova properties with varying resolution for the 20 $M_{\odot}$, 1/50 $Z_\odot$ rapidly rotating model ($V_{ini}$ = 600 km $\rm s^{-1}$). Models \texttt{20a} to \texttt{20e} represent  different mass and time resolutions, which in MESA are controlled by \textsf{mesh\_delta\_coeff} and \textsf{varcontrol\_target} leading to different number of zones ($N_{\rm zones}$) and steps ($N_{\rm steps}$) used in the simulation. When $\alpha_{\rm OV}$ = 0, convection overshooting is turned off. $\alpha_{\rm MLT}$ is the mixing length parameter. Key properties of the models are given, including the mass, radius, He core mass, CO core mass, Fe core mass and the compactness parameter $\xi_{2.5}$ at pre-collapse; besides, $\rm ^{12}C_{ign}$ is given at center C ignition.
The units of mass and radius are $M_\odot$ and $R_\odot$, respectively.}
\label{tab}

\end{deluxetable*}
\subsubsection{Explodability for $M_{4}$ and $\mu_{4}$}
\begin{figure}
  \includegraphics[width=\columnwidth]{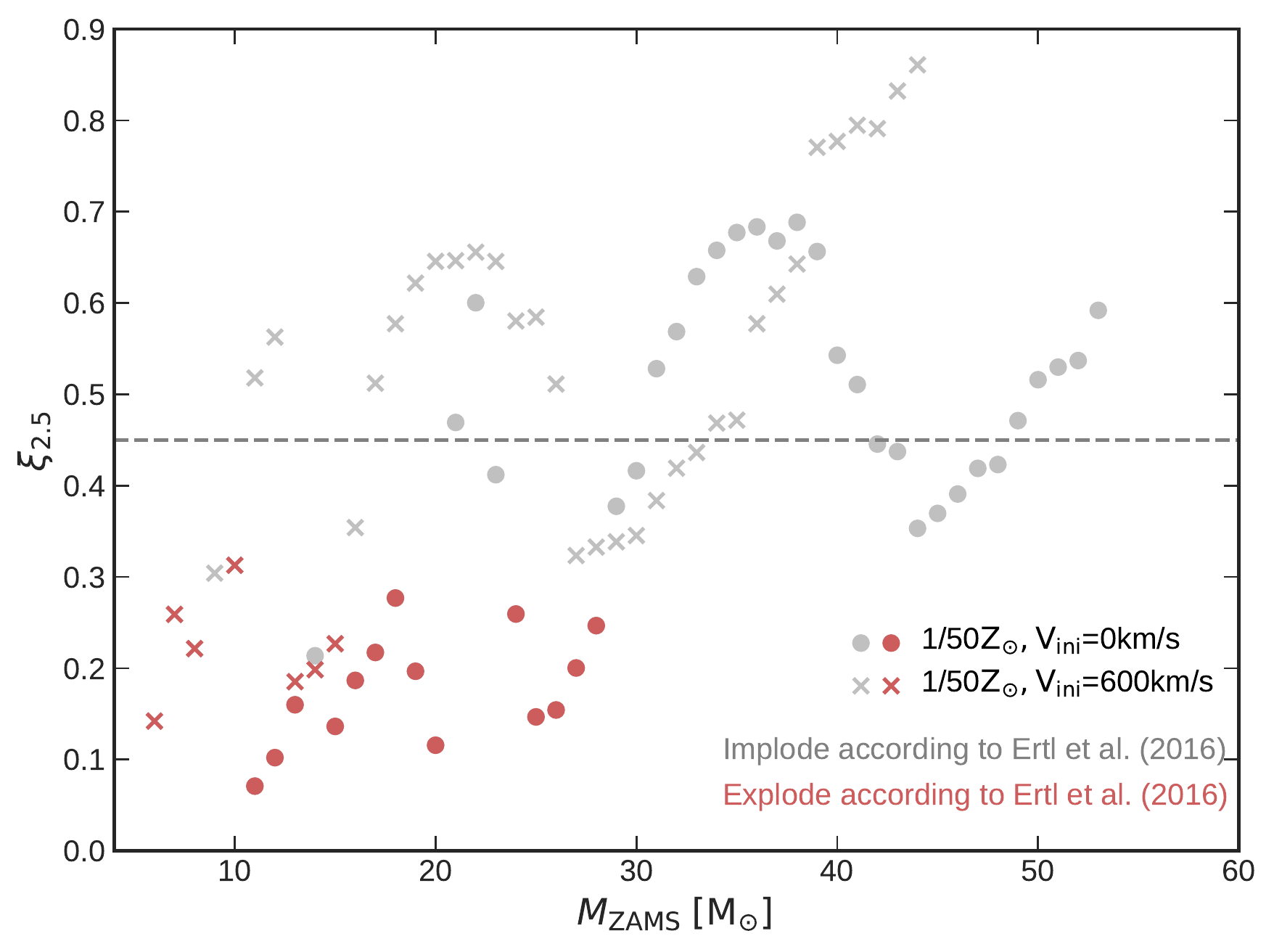}
  \caption{The explodability of models, as a function of their $M_{\rm ZAMS}$ for the criterion of \cite{Ertl2016}, which are given by different colorful marks.
  The red and gray marks present the explode and the implode models.
  The dotted line at $\xi_{2.5}$ = 0.45 separates models that might explode (below the line) or implode (above the line) according to \cite{Connor2011}.
  }
  \label{fig:ertl}
\end{figure}
Figure \ref{fig:ertl} shows that the effect of rotation on the trend of the explodability for the criterion of \cite{Ertl2016} is similar to $\xi_{2.5}$. The two-parameter criterion indicates that the explodability is also nonmonotonic with the initial mass. For nonrotating models, FSN island appears around $\sim$ 21 $M_{\odot}$. For rapidly rotating models, FSN island appears at lower $M_{\rm ZAMS}$ $\sim$ 11 $M_{\odot}$. In addition, based on Figure \ref{fig:ertl}, we note that the Ertl criterion yields fewer successful explosion models. $M_4$ is closely related to Si core mass, and $\mu_4$ is an almost completely congruent pattern with $\xi_{2.5}$ \citep{Sukhbold2020}. The main reason may be that the value of $\xi_{2.5}$ = 0.45 is slightly larger.
\subsection{Effects of Magnetic Field on Explodability}\label{sec:Metallicity}

\begin{figure}
  \centering
  \includegraphics[width=1.0\columnwidth]{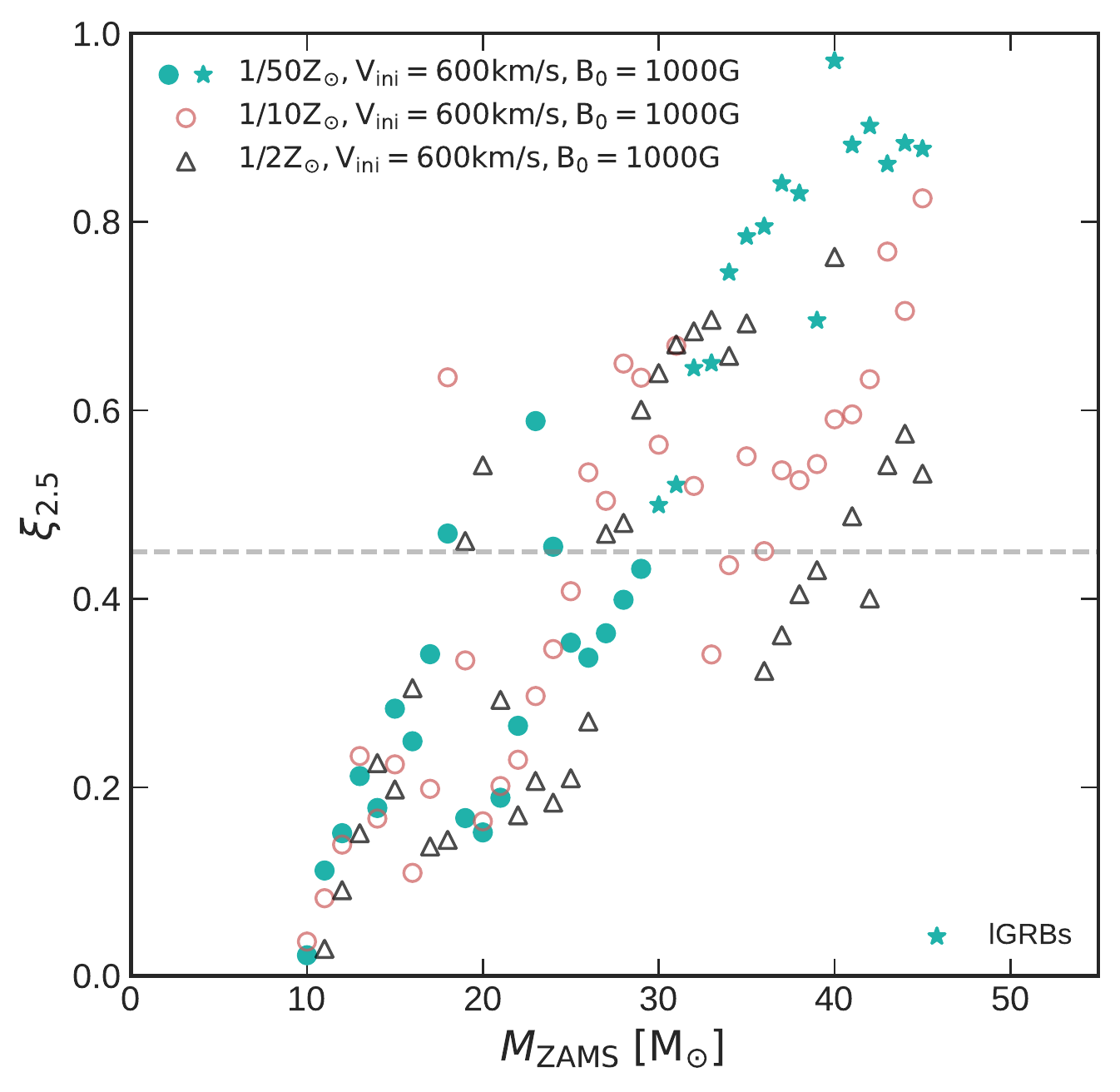}\\  
  \caption{Same as Figure \ref{fig:kexi} but for the 1/2 $Z_{\odot}$ models and the magnetic field models. For 1/50 $Z_{\odot}$ models with magnetic field and $V_{\rm ini}$ = 600 km $\rm s^{-1}$, only $M_{\rm ZAMS}$ $\ge$ 25 $M_{\odot}$ experience CHE.
  }\label{fig:kexi_mag}
\end{figure}
\begin{figure*}
  \centering
  \includegraphics[totalheight=5.2in,width=5.5in]{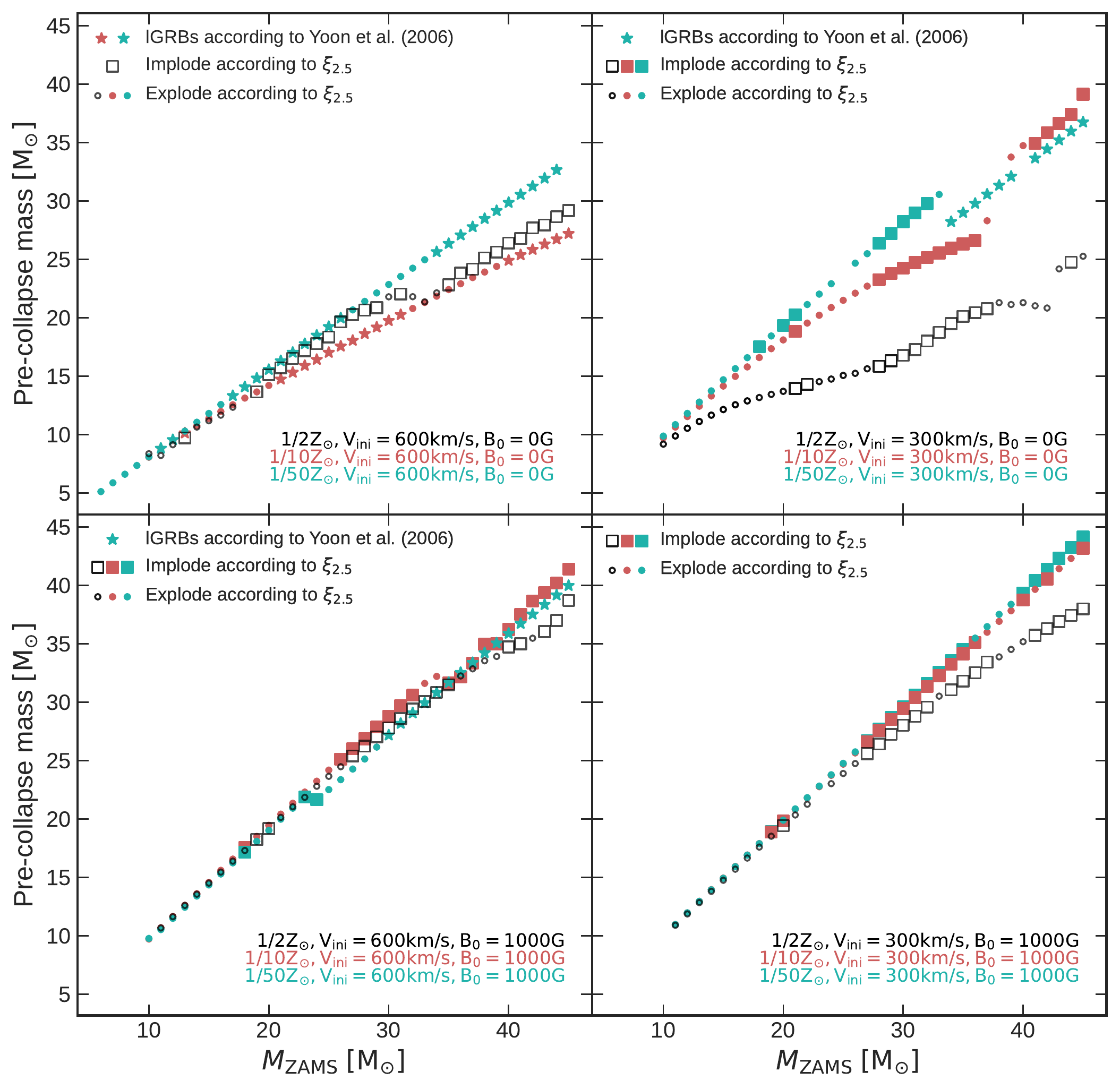}
  \caption{Final mass at core collapse as a function of initial mass of different initial models. The explosion (dotted markers) and implosion (square markers) are determined by $\xi_{2.5}<$ 0.45 or $\xi_{2.5}\ge$ 0.45 according to \cite{Connor2011}. LGRBs (star markers) are determined to be implosion models where $j_{\rm co}\ge j_{\rm Kerr, LSO}$ according to \cite{Yoon2006}.}.
	\label{fig:rem_mass}
\end{figure*}

As Figure \ref{fig:kexi_mag} shown, for the magnetic field model, FSN occurs at 18 $M_{\odot}$ and 23 $M_{\odot}$, and lGRBs occur only where $M_{\rm ZAMS}\ge$ 30 $M_{\odot}$ in the 1/50 $Z_{\odot}$ models, which is significantly different from the models without a magnetic field (Figure \ref{fig:kexi}).
The main reason is that the magnetic field reduces the stellar rotation via magnetic braking, which is unfavorable for lGRBs.
Simultaneously, the magnetic field constraints the mass-loss rates. Therefore, the rapid rotating massive stars with strong magnetic field
may evolve into the heavy-mass black holes ($M_{\rm BH}>\sim 30 M_\odot$) detected by LIGO \citep{2020PhRvL.125j1102A,2021arXiv210801045T}.

Figure \ref{fig:rem_mass} shows the pre-collapse masses of all rotating models with different metallicities and magnetic fields. In the models undergoing
lGRBs (marked by stars) or successful CCSN (marked by circles), the majority of the pre-collapse mass is ejected during SN explosion. These massive stars hardly form heavy-mass black holes, which is shown by the two top panels in Figure 10 except for the 1/10 $Z_{\odot}$ models with $V_{\rm init}=300$ km s$^{-1}$ and $M_{\rm ZAMS}>40$$M_{\odot}$. However, for the models with 1000 G magnetic field, the pre-collapse masses are higher than 30 $M_\odot$ when $M_{\rm ZAMS}>\sim35$ $M_\odot$. Most of them, except for the models with $V_{\rm init}=600$ km s$^{-1}$, will undergo FSN marked by squares, and form the heavy-mass black holes ($M_{\rm BH}>\sim 30 M_\odot$). Therefore, in our models, the heavy-mass black holes detected by LIGO may originate from the rapidly rotating massive stars with strong magnetic fields but not the ones with very low metallicity.

\section{Conclusions}\label{sec:Conclusions}
Using the MESA stellar evolution code, we investigate models of FSN and lGRBs for different rotation velocities, metallicity, and magnetic fields by the explodability criteria ($\xi_{2.5} \ge$ 0.45) and the lGRB generation criteria ($j_{\rm co}\ge j_{\rm Kerr, LSO}$).
We summarize the main conclusions as follows:

\begin{enumerate}
  \item Rapid rotation results in lower $\rm ^{12}C$ mass fraction when the central carbon is ignited, which makes $\xi_{2.5}\ge0.45$ for the models of rapidly rotating stars with  $M_{\rm ZAMS}$ $\sim$ 11 $M_{\odot}$. However, there are no islands of FSN around $\sim$ 11 $M_{\odot}$ since the angular momenta of these stars are very large so that $j_{\rm co}\ge j_{\rm Kerr, LSO}$ at the core collapse, and they evolve into lGRBs. Rapid rotation has complicated effects on $\xi_{2.5}$ via enhancing the stellar mass-loss rate and the chemical mixing efficiency, and it is unfavorable for FSN occurring but is conducive to lGRBs.
  \item The increase of metallicity does not change the island of FSN, but the inhibition of rotational mixing effect by high metallicity is not favorable for the producing of lGRBs. Low-metallicity stars undergo CHE to retain enough specific angular momentum to produce lGRBs.
  \item The fossil magnetic field can constrain the mass-loss rate even for rapid rotating stars, which results in the higher mass when there is a higher $\xi_{2.5}$.
  The magnetic braking triggered by the magnetic field can reduce the rotation velocity for high-metallicity models,
  which decreases $j_{\rm co}$ and is unfavorable for producing lGRBs. Therefore, in our simulations, 1/2 $Z_{\odot}$ and 1/10 $Z_{\odot}$ models with the magnetic field of 1000 G can experience FSNs and form black hole with a mass higher than $\sim$ 30 $M_\odot$ when their masses are higher than $\sim$ 35 $M_\odot$. Therefore, based on our models,  the heavy-mass black holes detected by LIGO may originate from the rapidly rotating massive stars with strong magnetic fields but not the ones with very low metallicity.
\end{enumerate}

\begin{acknowledgments}
We thank Dr. David Aguilera-Dena, for many helpful comments and suggestions.
This work received the generous support of the Nafional Natural Science Foundafion of Chinaunder grants U2031204,
12163005, and 12288102, the science research grants from the China Manned Space Project
with No. CMS-CSST-2021-A10,and the Xinjiang Uygur Autonomous Region Graduate Student Innovation Project
No. XJ2023G023, and the Natural Science Foundafion of Xinjiang Nos.
2021D01C075, No.2022D01D85,and 2022TSYCLJ0006.
\end{acknowledgments}
\bibliography{ref}
\bibliographystyle{aasjournal}

\end{document}